\documentclass[manuscript]{aastex}
\shorttitle{MHD modeling on geodesic grids}
\shortauthors{Florinski et al.}

\begin{document}

\title{MHD modeling of solar system processes on geodesic grids}

\author{V. Florinski\altaffilmark{1,2}, X. Guo\altaffilmark{2},
D. S. Balsara\altaffilmark{3}, and C. Meyer\altaffilmark{3}}

\altaffiltext{1}
{Department of Physics, University of Alabama, Huntsville, AL 35899}

\altaffiltext{2}
{Center for Space Plasma and Aeronomic Research, University of Alabama,
Huntsville, AL 35899}

\altaffiltext{3}
{Department of Physics, University of Notre Dame, Notre Dame, IN 46556}

\begin{abstract}
This report describes a new magnetohydrodynamic numerical model based on a
hexagonal spherical geodesic grid.
The model is designed to simulate astrophysical flows of partially ionized
plasmas around a central compact object, such as a star or a planet with a
magnetic field.
The geodesic grid, produced by a recursive subdivision of a base platonic solid
(an icosahedron), is free from control volume singularities inherent in
spherical polar grids.
Multiple populations of plasma and neutral particles, coupled via
charge-exchange interactions, can be simulated simultaneously with this model.
Our numerical scheme uses piecewise linear reconstruction on a surface of a
sphere in a local two-dimensional ``Cartesian'' frame.
The code employs HLL-type approximate Riemann solvers and includes facilities to
control the divergence of magnetic field and maintain pressure positivity.
Several test solutions are discussed, including a problem of an interaction
between the solar wind and the local interstellar medium, and a simulation of
Earth's magnetosphere.
\end{abstract}

\keywords{magnetohydrodynamics (MHD) --- methods: numerical --- 
planets and satellites: magnetic fields --- solar wind}

\section{Introduction}
Many astrophysical plasma processes occur is regions of space surrounding a
central compact object such as a star or a planet.
Examples include stellar winds, planetary magnetospheres, supernova blast waves,
and mass accretion onto compact objects.
In all these environments the central body (a star or a planet) is typically
much smaller than the characteristic scales of the plasma flows.
In solving this class of problem on a computer, radial grids are commonly
used because resolution can be readily increased near the origin.
The simplest and the most commonly used is the standard spherical polar
($r$, $\theta$, $\varphi$) grid \citep[e.g.,][]{washimi96, pogorelov98,
ratkiewicz98}.
This grid has a singularity on the $z$ axis, where the control volume
$\Delta V=r^2\Delta r\sin\theta\Delta\theta\Delta\varphi$, $\Delta r$,
$\Delta\theta$ and $\Delta\varphi$ being the grid cell dimensions in the radial,
latitudinal, and azimuthal directions, respectively, vanishes as
$\sin\theta\to 0$.
For explicit methods, this requires a small global time step to satisfy the
Courant stability condition for a system of hyperbolic conservation laws for the
entire grid (implicit or semi-implicit methods \citep[e.g.,][]{toth98} don't
suffer from this limitation).

Spherical grids are also used in simulating the transport of energetic charged
particle, such as galactic cosmic rays, in turbulent astrophysical flows
\citep{florinski09}.
Transport models based on stochastic trajectory (Monte-Carlo) methods also
suffer from the singularity on the polar axis.
For example, in modeling cosmic-ray transport in the heliosphere it is common to
align the $z$ axis with the solar rotation axis.
Because energetic particle transport (diffusion and drift) is very rapid at high
latitudes due to a weaker magnetic field, a model must take vanishingly small
time steps when a particle ventures close to the polar axis, which results in an
inferior overall model efficiency.

Time step requirements can be relaxed substantially by employing a grid that has
a (nearly) uniform solid angle coverage.
Examples include triangle, hexagon, or diamond based geodesic grids
\citep{du03, yeh07, upadhyaya10}, obtained by a recursive division of a base
platonic solid, and cubed sphere grids \citep{ronchi96, putman07}.
This paper introduces a framework for finite volume methods of solution of
hyperbolic conservation laws in three dimensions, such as gas-dynamic or MHD
systems, using spherical geodesic grids composed of hexagonal prism elements.
Results are illustrated on a three-dimensional simulation of solar rotation and
formation of corotating interaction regions (CIRs), an interaction between the
solar wind and the surrounding local interstellar medium or LISM, and a
simulation of Earth's magnetosphere.
The new framework can be employed to model a broad range of large-scale 3D
astrophysical plasma flows around a compact object where high computational
efficiency is a priority.

\section{Grid structure}
Our three-dimensional grid consist of a 2D geodesic unstructured grid on a
sphere combined with a concentric nonuniform radial stepping with smaller cells
near the origin.
The 2D surface grid is a Voronoi tesselation of a sphere produced from a dual
triangular (Delaunay) tesselation.
The latter is generated by a recursive subdivision of an icosahedron.
We use the geodesic grid generator software developed by the ICON project
(\textit{http://icon.enes.org}) for use in atmospheric circulation modeling.
An optimization algorithm \citep{heikes95}, included in their code, produces a
mesh with a difference in spherical surface areas between the largest and the
smallest cells of less than 10\%.

\begin{figure}
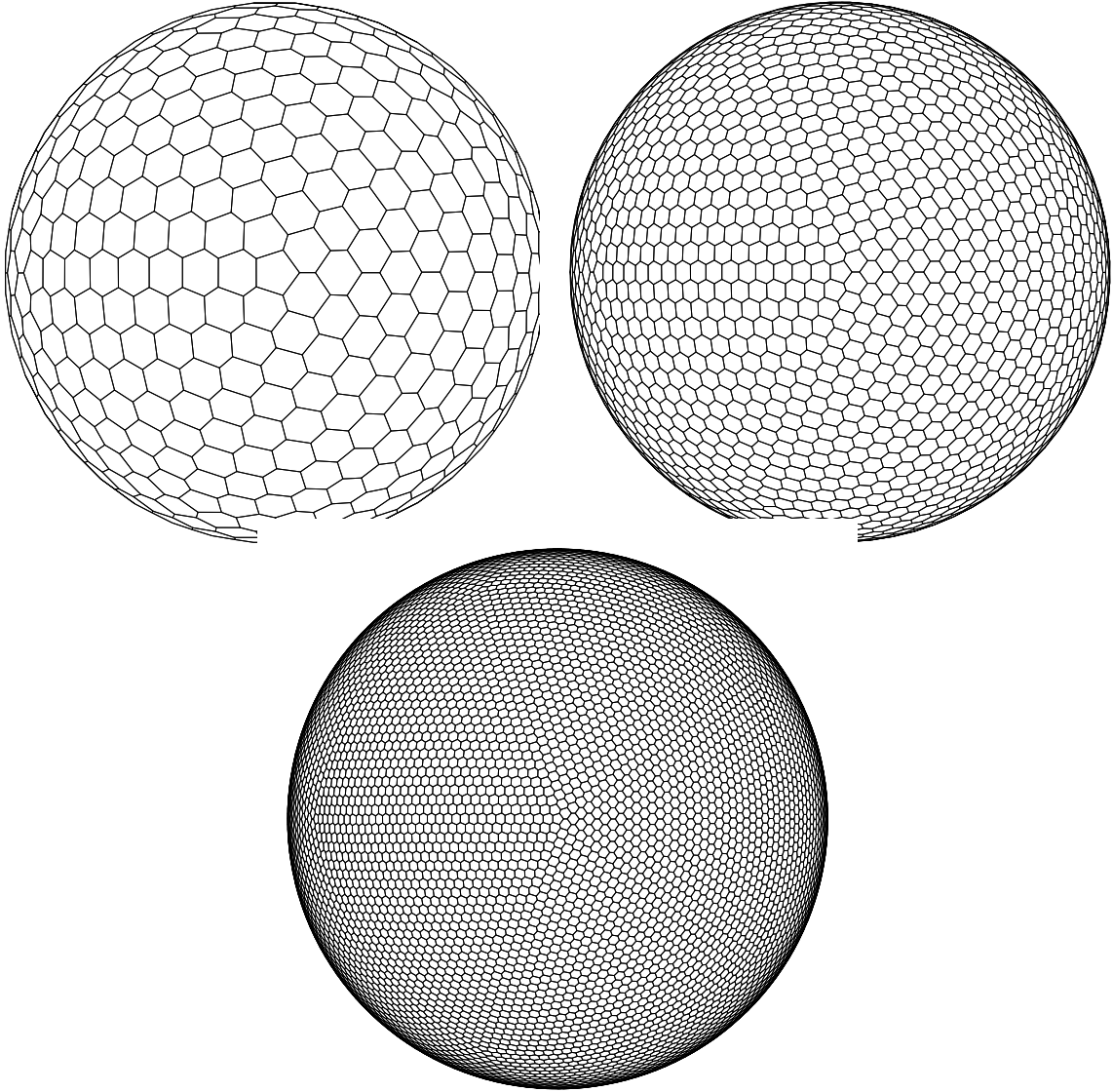

\epsscale{0.45}
\plotone{level4.eps}
\plotone{level5.eps}
\plotone{level6.eps}
\caption{From left to right: level 3, 4, and 5 geodesic Voronoi grids produced
by a recursive division of the base icosahedron.}
\label{fig_geogrids}
\end{figure}

The number of vertices (V), edges (E) and faces (F) on a grid produced by
$l$th division is given by
\begin{equation}
N_\mathrm{V}=20\cdot 2^{2l},\quad N_\mathrm{E}=30\cdot 2^{2l},
\quad N_\mathrm{F}=10\cdot 2^{2l}+2.
\end{equation}
In this notation a level 0 grid is dual to the original icosahedron projected
onto a unit sphere.
The base shape of a control volume in a finite volume method is a spherical
hexagonal prism with the exception of 12 pentagonal prisms located at the
vertices of the original icosahedron.
Level 3, 4, and 5 hexagonal geodesic grids are shown in Figure \ref{fig_geogrids}.

\begin{figure}
\epsscale{0.8}
\plotone{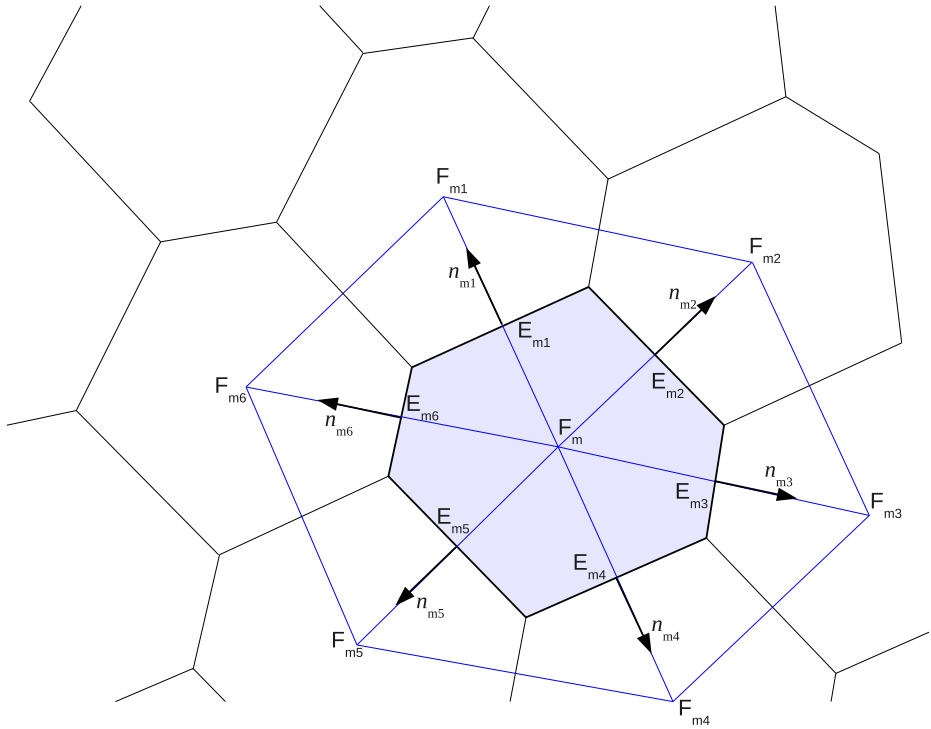}
\caption{A close up view of the geodesic grid illustrating the relationships
between a face $\mathrm{F}_m$ and its neighboring faces
$\mathrm{F}_{m1}..\mathrm{F}_{m6}$.
Unit vectors normal to the edges of the Voronoi face
$\hat\mathbf{n}_{m1}..\hat\mathbf{n}_{m6}$ are shown.
A fragment of the Delaunay grid is drawn with blue lines.}
\label{fig_faces}
\end{figure}

A more detailed view of the Voronoi grid structure is given in Figure
\ref{fig_faces}.
A hexagonal face $\mathrm{F}_m$ (shaded) is shown surrounded by six adjacent
faces $\mathrm{F}_{m1}..\mathrm{F}_{m6}$.
The face centers of the dual triangle-based Delaunay grid (blue lines) are
located at the Voronoi vertices (V); likewise, the former's vertices are at
the Voronoi grid's face centers (F).
The edges of the Voronoi and Delaunay grids on a sphere are mutually orthogonal
and intersect at their midpoints.
To achieve acceptable resolution in typical astrophysical flow modeling problems
level 5 or higher geodesic grids should be used.
Most of our simulations use level 6 mesh containing 40,962 Voronoi polygons.
We emphasize that these 40,962 hexagons are distributed evenly over the surface
of each spherical layer of cells.
This gives us a resolution of about $3\times 10^{-4}$ steradians in solid angle
which corresponds to an angular resolution of one degree.
By condensing the mesh in the radial direction, one can achieve a further degree
of refinement as required by the problem.

We introduce a set of unit vectors normal to the edges of the Voronoi grid
$\hat\mathbf{n}_{mn}$, where the index $m$ refers to the $m$th face and
$n=[1,6]$ is the number of the edge counted in a counter-clockwise direction.
The unit vectors are tangential to the surface of the sphere.
The corresponding edge lengths, measured along great circles, are designated
$L_{mn}$.
The outward radial unit vector at the cell center is $\hat\mathbf{r}_{m}$.
We also designate the area of the face on the unit sphere as $A_m$.
In this notation the control volume is equal to
\begin{equation}
\Delta V_{im}=A_{m}r_i^2\Delta r_i,
\end{equation}
where $i$ is the index on the radial axis.
The areas $A_m$ are calculated by dividing each hexagonal face into six (five
for pentagons) spherical triangles and adding up their areas using standard
expressions from spherical trigonometry.
Having defined our cell dimensions is this way we can proceed to integrate a
system of conservation laws inside a control volume.

It is worth pointing out that a similar geodesic-mesh-based model was developed
by \citet{nakamizo09} 
Their mesh was generated from a dodecahedron by first dividing each face into
five triangles followed by a recursive subdivision of each triangle into four
smaller triangles.
The resulting unstructured grid topology is similar (but not identical) to our
dual Delaunay grid.
Interestingly, in the model of \citet{nakamizo09} computations are also
performed on a hexagonal grid, generated by connecting the centroids of the
Delaunay triangles.

\section{MHD conservation laws}
For the heliospheric and magnetospheric problems (\S 6) we solve a modified set
of MHD equations, written in terms of conservative variables $\mathbf{U}$ and
fluxes $\mathbf{F}$ as
\begin{equation}
\frac{\partial\mathbf{U}}{\partial t}+\nabla\cdot\mathbf{F}=\mathbf{Q},
\end{equation}
where
\begin{equation}
\mathbf{U}=\left(\begin{array}{c}
\rho \\
\rho\mathbf{u} \\
e \\
\mathbf{B}
\end{array}\right),\quad
\mathbf{F}=\left(\begin{array}{c}
\rho\mathbf{u} \\
\rho\mathbf{u}\mathbf{u}+p\mathbf{I}-\mathbf{B}\mathbf{B}/(4\pi) \\
(e+p)\mathbf{u}-\mathbf{B}(\mathbf{u}\cdot\mathbf{B})/(4\pi) \\
\mathbf{u}\mathbf{B}-\mathbf{B}\mathbf{u}
\end{array}\right)
\end{equation}
in CGS units.
Here $\rho$ is density, $\mathbf{u}$ is velocity, $\mathbf{B}$ is magnetic
field, $\mathbf{I}$ is a unit dyadic, $p=p_\mathrm{g}+B^2/(8\pi)$ is
the total pressure, $p_\mathrm{g}$ being the gas kinetic pressure, and the
energy density $e$ is given by
\begin{equation}
e=\frac{\rho u^2}{2}+\frac{p_\mathrm{g}}{\gamma-1}
+\frac{B^2}{8\pi}.
\end{equation}

We employ two alternative models to control the divergence of magnetic field.
The first is the numerical scheme proposed by \citet{powell99}, where numerical
magnetic field divergence is advected out of the simulation domain with the flow
velocity.
This scheme modifies the system of conservation laws with a hyperbolic source
term
\begin{equation}
\mathbf{Q}=-\nabla\cdot\mathbf{B}\left(\begin{array}{c}
0 \\
\mathbf{B}/(4\pi) \\
\mathbf{u}\cdot\mathbf{B}/(4\pi) \\
\mathbf{u}
\end{array}\right).
\end{equation}
The second scheme employs a generalized Lagrange multiplier (GLM) $\psi$ for a
mixed hyperbolic-parabolic correction \citep{dedner02}.
The system (4) is extended with an additional transport equation for $\psi$
\begin{equation}
\frac{\partial\psi}{\partial t}+\nabla\cdot(c_h^2\mathbf{B})=-\frac{c_h^2}{c_p^2}\psi,
\end{equation}
where $c_h$ is a constant, isotropic advection speed, taken to be somewhat
faster than the fastest wave speed in the problem, and $c_p$ is related to the
rate of decay of $\psi$.
\citet{dedner02} proposed two methods to fix the value of $c_p$: (a) by fixing
the time rate of decay of the GLM variable $r_d=\exp(-\Delta t c_h^2/c_p^2)$,
where $\Delta t$ is the time step, and (b) by fixing the characteristic length
over which the decay occurs, given by $l_d=c_p^2/c_h$.
Both methods are available in our code.
In the GLM scheme the conservation law for magnetic field (Faraday's law) is
modified to read
\begin{equation}
\frac{\partial\mathbf{B}}{\partial t}
+\nabla\cdot(\mathbf{u}\mathbf{B}-\mathbf{B}\mathbf{u}+\psi\mathbf{I})=0.
\end{equation}

\begin{figure}
\epsscale{0.7}
\plotone{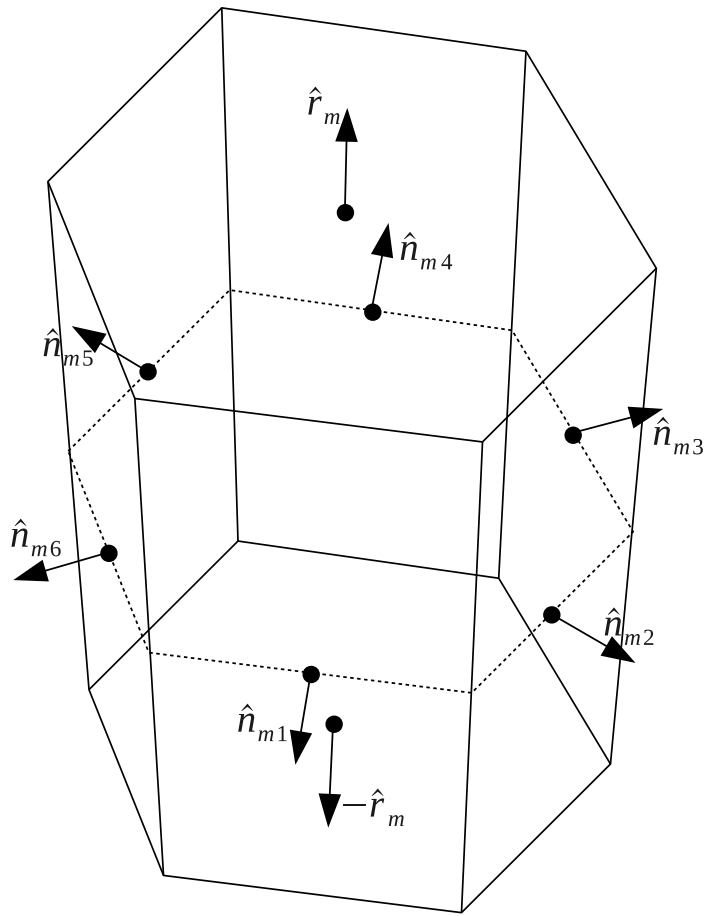}
\caption{A prismatic control volume showing the unit vectors normal to the
interfaces.}
\label{fig_convol}
\end{figure}

The system (3) is integrated over a control volume $\Delta V_{im}$ shown in
Figure \ref{fig_convol} to obtain the finite volume method
\begin{eqnarray}
\Delta V_{im}\frac{\Delta\mathbf{U}_{im}}{\Delta t}
=-A_{m}\left(r_{i+1/2}^2\mathbf{F}_{i+1/2,m}
-r_{i-1/2}^2\mathbf{F}_{i-1/2,m}\right)\cdot\hat\mathbf{r}_m \nonumber \\
-r_i\Delta r_i\sum_{n=1}^6 L_{mn}\mathbf{F}_{i(mn)}\cdot\hat\mathbf{n}_{mn}
+\Delta V_{im}\mathbf{Q}_{im}.
\end{eqnarray}
Here $\mathbf{F}_{i(mn)}$ is the flux at the center of the edge shared by the
$m$th cell and its $n$th neighbor (where $n=[1,6]$).
Note that the source term $\mathbf{Q}$ does not involve the right hand side of
Eq. (7).
The parabolic correction is applied by multiplying the value of $\psi$ obtained
from the finite volume scheme (9) by the decay factor,
\begin{equation}
\psi\to\psi\times\left\lbrace
\begin{array}{ll}
r_d, & \mathrm{method\;(a)}, \\
e^{-\Delta t c_h/l_d}, & \mathrm{method\;(b)}.
\end{array}\right.
\end{equation}
In our simulations we typically use $0.9<r_d<1$ and $l_d$ equal to several
times the smallest linear grid size.

The divergence of the magnetic field is obtained from Gauss's theorem in the
same way as $\nabla\cdot\mathbf{F}$ is calculated in Eq. (9).
More generally, the divergence and curl operators acting on an arbitrary vector
$\mathbf{v}$ may be written as
\begin{equation}
\nabla\cdot\mathbf{v}=\frac{\hat\mathbf{r}_m\cdot(r_{i+1/2}^2\mathbf{v}_{i+1/2,m}
-r_{i-1/2}^2\mathbf{v}_{i-1/2,m})}{r_i^2\Delta r_i}
+\sum_{n=1}^6\frac{L_{mn}\hat\mathbf{n}_{mn}\cdot\mathbf{v}_{i(mn)}}{r_iA_{m}},
\end{equation}
\begin{equation}
\nabla\times\mathbf{v}=\frac{\hat\mathbf{r}_m\times(r_{i+1/2}^2\mathbf{v}_{i+1/2,m}
-r_{i-1/2}^2\mathbf{v}_{i-1/2,m})}{r_i^2\Delta r_i}
+\sum_{n=1}^6\frac{L_{mn}\hat\mathbf{n}_{mn}\times\mathbf{v}_{i(mn)}}{r_iA_{m}}.
\end{equation}
An evaluation of a curl is necessary when modeling energetic charged particle
transport, where the particle's drift velocity is proportional to
$\nabla\times(\mathbf{B}/B^2)$.
The values of primitive variables at face centers $\mathbf{v}_{i\pm 1/2}$ and
$\mathbf{v}_{i(mn)}$ may be approximated as arithmetic averages of the values in
the two cells separated by the face.
A more accurate approach, adopted here, is to use interface resolved states
obtained from a solution to the corresponding Riemann problem (see \S5).

The finite volume system of conservation laws is integrated with a second order
unsplit TVD-like method (see below).
Right and left interface values are calculated in the usual way using some
appropriate linear reconstruction to achieve second-order spatial accuracy.
Fluxes are calculated from a solution to a one-dimensional (projected) Riemann
problem at each cell interface.
Finally, time is advanced using either a first order (Euler) or, more commonly,
a second order (Runge-Kutta) scheme, depending on the nature of the problem.

\section{Reconstruction}
To achieve second order spatial accuracy we employ limited piecewise linear
reconstruction on primitive variables $\mathbf{V}=(\rho,\mathbf{u},p_\mathrm{g},
\mathbf{B}, \psi)^T$.
In the radial direction the simplest and the most robust limiter available is
the MinMod, with slopes $\mathbf{S}_{im}$ given by
\begin{equation}
\mathbf{S}^\mathrm{MM}_{im}=\mathrm{minmod}(\mathbf{S}^-_{im}, \mathbf{S}^+_{im}),
\end{equation}
where the left and the right slopes on an asymmetric stencil are, respectively
\begin{equation}
\mathbf{S}^-_{im}=2\frac{\mathbf{V}_{im}-\mathbf{V}_{i-1,m}}
{\Delta r_{i-1}+\Delta r_i},\quad
\mathbf{S}^+_{im}=2\frac{\mathbf{V}_{i+1,m}-\mathbf{V}_{im}}
{\Delta r_i+\Delta r_{i+1}}.
\end{equation}
Also available is the more compressive monotonized central (MC) limiter
\citep{vanleer77} with
\begin{equation}
\mathbf{S}^\mathrm{MC}_{im}=\mathrm{minmod}\left[2\mathbf{S}^-_{im},2\mathbf{S}^+_{im},
\frac{(\Delta r_i+\Delta r_{i+1})\mathbf{S}^-_{im}
+(\Delta r_{i-1}+\Delta r_i)\mathbf{S}^+_{im}}
{\Delta r_{i-1}+2\Delta r_i+\Delta r_{i+1}}\right].
\end{equation}
The third option is the weighted essentially non-oscillatory (WENO) limiter
\begin{equation}
\mathbf{S}^\mathrm{WENO}_{im}=\bar\mathbf{w}^-_{im}\mathbf{S}^-_{im}
+\bar\mathbf{w}^+_{im}\mathbf{S}^+_{im},
\end{equation}
where the WENO weights $\bar\mathbf{w}$ are given by
\begin{equation}
\bar\mathbf{w}^-_{im}=\frac{({\mathbf{S}^-_{im}}^2+\epsilon)^{-p}}
{({\mathbf{S}^-_{im}}^2+\epsilon)^{-p}+({\mathbf{S}^+_{im}}^2+\epsilon)^{-p}},
\quad
\bar\mathbf{w}^+_{im}=\frac{({\mathbf{S}^+_{im}}^2+\epsilon)^{-p}}
{({\mathbf{S}^-_{im}}^2+\epsilon)^{-p}+({\mathbf{S}^+_{im}}^2+\epsilon)^{-p}},
\end{equation}
where $p$ is an integer constant here taken to be 2, and $\epsilon$ is a small
number, which we took to be $10^{-12}$ in our simulations.
The $(\mathbf{S}^+_{im})^{2p}$ and $(\mathbf{S}^-_{im})^{2p}$ terms are
traditionally referred to as smoothness measures in WENO methodology.

\begin{figure}
\epsscale{0.7}
\plotone{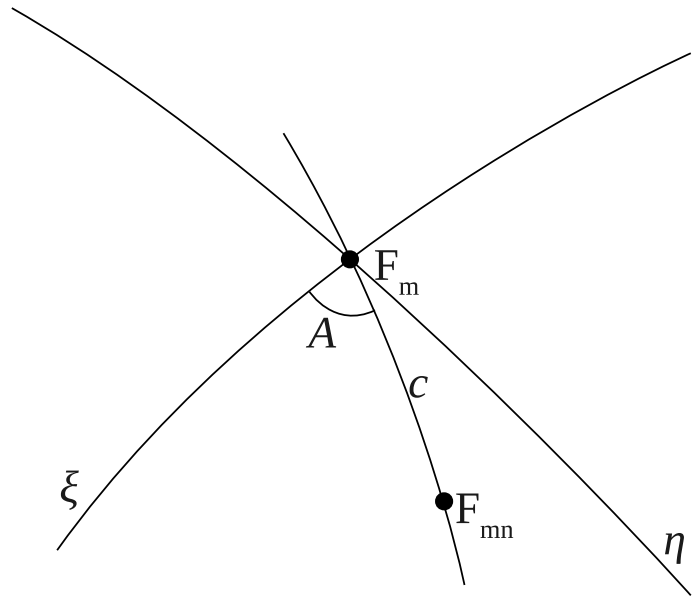}
\caption{Calculation of adjacent cell center (F$_{mn}$) coordinates in a local
coordinate frame associated with a face F$_m$.
Every line shown is a segment of a great circle.}
\label{fig_sphplan}
\end{figure}

For two-dimensional reconstruction on the surface of a sphere the code can use
either a minimum angle plane (MAPR) method \citep{christov08} or a 2D version of
the weighted essentially non-oscillatory (WENO) scheme \citep{friedrich98}.
Common to both schemes, a local two-dimensional coordinate system ($\xi$,$\eta$)
is introduced on the sphere, with its origin at the face center F$_m$.
The coordinates of the six adjacent cell centers ($\xi_{mn}$, $\eta_{mn}$) are
then calculated in this frame.
These coordinates are measured along two arbitrary great circles intersecting at
right angles at the position of the central face F$_m$.
The procedure is illustrated in Figure \ref{fig_sphplan}.
The angle $A$ and the great circle distance between the face centers $c$ are
effectively polar coordinates on the surface of a sphere.
The local coordinates of F$_{mn}$ are calculated as
\begin{equation}
\xi_{mn}=c\cos A,\quad\eta_{mn}=c\sin A.
\end{equation}

Next, the six two-dimensional slopes $\mathbf{S}^\xi$, $\mathbf{S}^\eta$ are
calculated from the triangles with vertices located at the cell centers F$_m$,
F$_{mn}$, F$_{m,n+1}$ (shown in blue in Figure \ref{fig_faces}) as
\begin{equation}
\mathbf{S}^\xi_{imn}=\frac{\eta_{mn}(\mathbf{V}_{im,n+1}-\mathbf{V}_{im})
-\eta_{m,n+1}(\mathbf{V}_{imn}-\mathbf{V}_{im})}
{\eta_{mn}\xi_{m,n+1}-\xi_{mn}\eta_{m,n+1}},
\end{equation}
\begin{equation}
\mathbf{S}^\eta_{imn}=-\frac{\xi_{mn}(\mathbf{V}_{im,n+1}-\mathbf{V}_{im})
-\xi_{m,n+1}(\mathbf{V}_{imn}-\mathbf{V}_{im})}
{\eta_{mn}\xi_{m,n+1}-\xi_{mn}\eta_{m,n+1}}.
\end{equation}
In the MAPR method we evaluated the average slopes as
\begin{equation}
\bar\mathbf{S}_{imn}=\sqrt{{\mathbf{S}^\xi_{imn}}^2+{\mathbf{S}^\eta_{imn}}^2}.
\end{equation}
The reconstructed slopes $\mathbf{S}^{\mathrm{MAPR}\,\xi,\eta}_{im}$ are those
given by Eqs. (19) and (20) for which the average slope (20) is the smallest.
Thus the method is a two-dimensional equivalent of the MinMod limiter.
In the WENO method the reconstructed slopes are weighted arithmetic averages of
all six slopes, namely
\begin{equation}
\mathbf{S}^{\mathrm{WENO}\,\xi,\eta}_{im}=\sum_{n=1}^6
\bar\mathbf{w}_{imn}\mathbf{S}^{\xi,\eta}_{imn}.
\end{equation}
The weights $\bar\mathbf{w}_{imn}$ of each slope are calculated as
\begin{equation}
\bar\mathbf{w}_{imn}=\frac{\mathbf{w}_{imn}}{\sum_n\mathbf{w}_{imn}},
\end{equation}
where
\begin{equation}
\mathbf{w}_{imn}=\left({\mathbf{S}^\xi_{imn}}^2+{\mathbf{S}^\eta_{imn}}^2
+\epsilon\right)^{-p},
\end{equation}
where we again use $p=2$ and $\epsilon=10^{-12}$.
Because edge centers lie midway between the corresponding two face centers
F$_m$ and F$_{mn}$ on the Voronoi grid, the reconstructed values at edge
midpoints $\mathbf{V}_{i(mn)}$ can be computed as
\begin{equation}
\mathbf{V}_{i(mn)}=\mathbf{V}_{im}+\frac{1}{2}\mathbf{S}^\xi_{im}\xi_{mn}
+\frac{1}{2}\mathbf{S}^\eta_{im}\eta_{mn}.
\end{equation}
These values are used to calculate the intercell fluxes according to Eqs. (4),
(7), and (8).

The code also implements a slope flattening algorithm that reduces the value of
the slopes calculated by the reconstruction module in the vicinity of strong
compressions (shocks).
This prevents the occurrence of oscillations downstream of the shock.
To construct a flattener, we calculate the minimum value of the fast
magnetosonic wave speed $a^\mathrm{min}_{f,im}$ in each computational cell and
its neighbors in the same spherical layer and in the layers above and below (a
total of 21 cells).
The shock detector function in each cell $d_{im}$ is then calculated as
\citep{balsara09}
\begin{equation}
d_{im}=\mathrm{min}\left[1,\left|\frac{(\nabla\cdot\mathbf{u})_{im}\Delta l_{im}}
{a^\mathrm{min}_{f,im}\delta }+1\right|\right]
H\left[-\frac{(\nabla\cdot\mathbf{u})_{im}\Delta l_{im}}
{a^\mathrm{min}_{f,im}\delta }-1\right],
\end{equation}
where $\Delta l_{im}$ is a characteristic dimension of the cell $\Delta V_{im}$,
$\delta$ is a constant of order 1 and $H$ is the Heaviside step function.
Subsequently, the slope in the cell $im$ is calculated as a weighted sum of
a slope obtained with the standard limiter, such as WENO, and that from a more
diffusive limiter, such as MinMod or MAPR.
For example, in the radial direction we could use
\begin{equation}
\mathbf{S}_{im}=(1-d_{im})\mathbf{S}^\mathrm{WENO}_{im}
+d_{im}\mathbf{S}^\mathrm{MM}_{im}.
\end{equation}

\section{Riemann solvers}
The fluxes $\mathbf{F}$ are calculated from an (approximate) solution to the
one-dimensional Riemann problem at each interface between the prismatic cells.
A suitable solver must be able to handle supersonic and transonic flows without
losing positivity.
Our tests revealed that modern HLL-type solvers \citep{batten97,gurski04}
were generally superior to other solver types for the solar wind-LISM
interaction problem, where they were least likely to produce a negative pressure
upstream of a very strong (Mach number $>10$) shock.
Genuinely multi-dimensional Riemann solvers are now appearing in the literature
\citep{balsara10, balsara12}, and they offer substantial advantages on logically
rectangular meshes.
However, the analogous work for unstructured meshes is the topic of vigorous
research and was not incorporated in the present work.

An HLLC solver consists of four states: the left and the right unperturbed
states plus two intermediate states separated by a tangential discontinuity.
Designating the left and the right bounding wave speeds of the Riemann fan by
$S_l$ and $S_r$, respectively, the intercell flux may be written as
\begin{equation}
\mathbf{F}=\left\lbrace\begin{array}{ll}
\mathbf{F}_l, & S_l>0, \\
\mathbf{F}_l+S_l(\mathbf{U}^*_l-\mathbf{U}_l), & S_l\leq 0\leq S^*, \\
\mathbf{F}_r+S_r(\mathbf{U}^*_r-\mathbf{U}_r), & S^*\leq 0\leq S_r, \\
\mathbf{F}_r, & S_r<0,
\end{array}\right.
\end{equation}
where $\mathbf{F}_l=\mathbf{F}(\mathbf{U}_l)$, 
$\mathbf{F}_r=\mathbf{F}(\mathbf{U}_r)$, and $S^*$ is the speed of the
intermediate wave (a tangential discontinuity).
Because in a HLLC solver the normal velocity component and the total pressure
only change across the outermost waves, the speed of the tangential
discontinuity is readily calculated by applying the Rankine-Hugoniot conditions
across these waves.
This yields the speed
\begin{equation}
S^*=\frac{\rho_r u_{n,r}(S_r-u_{n,r})-p_r+B_{n,r}^2/(4\pi)
-\rho_l u_{n,l}(S_l-u_{n,l})+p_l-B_{n,l}^2/(4\pi)}
{\rho_r(S_r-u_{n,r})-\rho_l(S_l-u_{n,l})},
\end{equation}
where $u_{n(l,r)}$ and $B_{n(l,r)}$ are the normal-projected velocities and
magnetic fields in the left and right states, respectively.
Suppose two prismatic cells $(i,m_1)$ and $(i,m_2)$ share an interface with an
index $n_1=[1,6]$ in the neighbor list of the first cell and $n_2=[1,6]$ in
the neighbor list of the second cell (the definitions of ``first'' and
'`second'' are arbitrary here; they could be defined, for example, by using the
condition that $m_1<m_2$).
Then a normal velocity projection is defined as
\begin{equation}
u_{n,l}=\mathbf{u}_{i(m_1n_1)}\cdot\hat\mathbf{n}_{m_1n_1},\quad
u_{n,r}=\mathbf{u}_{i(m_2n_2)}\cdot\hat\mathbf{n}_{m_1n_1},
\end{equation}
where $\mathbf{u}_{i(m_1n_1)}$ and $\mathbf{u}_{i(m_2n_2)}$ are the
reconstructed velocities given by (25), and $\hat\mathbf{n}_{n_1n_1}$ is the
unit vector normal to the interface $n_1$ of the cell $(i,m_1)$, pointing
outward.
The values for $B_{n(l,r)}$ are computed in the same way.

Several HLLC MHD solvers may be found in the literature, distinguished by their
choice of the tangential velocity and magnetic field components in the
intermediate states $\mathbf{U}_{l,r}^*$ (unlike in gas dynamics, these states
are not unique in MHD).
Currently we employ a solver proposed by \citet{li05}.
Its main feature is that no jump in magnetic field is permitted across the
tangential discontinuity.

A second option available in our model is the HLLD Riemann solver
\citep{miyoshi05}.
This type of solver incorporates two additional states $\mathbf{U}_{l,r}^{**}$
separated from the corresponding ``single star'' states by rotational
(Alfv\'enic) discontinuities, propagating to the left and to the right of the
middle wave with speeds $S_l^*$ and $S_r^*$ respectively, given by
\begin{equation}
S^*_l=S^*-\frac{|B^*_n|}{(4\pi\rho^*_l)^{1/2}},\quad
S^*_r=S^*+\frac{|B^*_n|}{(4\pi\rho^*_r)^{1/2}},
\end{equation}
where $B^*_n$ is given by Eq. (34) below, and
\begin{equation}
\rho^*_l=\rho_l\frac{S_l-u_{n,l}}{S_l-S^*},\quad
\rho^*_r=\rho_r\frac{S_r-u_{n,r}}{S_r-S^*}.
\end{equation}
The intercell flux is computed as
\begin{equation}
\mathbf{F}=\left\lbrace\begin{array}{ll}
\mathbf{F}_l, & S_l>0, \\
\mathbf{F}_l+S_l(\mathbf{U}^*_l-\mathbf{U}_l), & S_l\leq 0\leq S^*_l, \\
\mathbf{F}_l+S_l(\mathbf{U}^*_l-\mathbf{U}_l)
+S^*_l(\mathbf{U}^{**}_l-\mathbf{U}^*_l), & S^*_l\leq 0\leq S^*, \\
\mathbf{F}_r+S_r(\mathbf{U}^*_r-\mathbf{U}_r)
+S^*_r(\mathbf{U}^{**}_r-\mathbf{U}^*_r), & S^*\leq 0\leq S^*_r, \\
\mathbf{F}_r+S_r(\mathbf{U}^*_r-\mathbf{U}_r), & S^*_r\leq 0\leq S_r, \\
\mathbf{F}_r, & S_r<0,
\end{array}\right.
\end{equation}
The HLLD solver is somewhat less robust than the HLLC counterpart because it has
a singularity when one of the extremal waves is a switch-on shock.
When this condition is encountered, the program falls back to the HLLC algorithm
which is singularity-free.

Because of nonlinearity of the solvers, under rare circumstances one of the
intermediate waves could fall outside of the bounding (fast) waves.
To prevent this from happening, we take the speed of the bounding waves to be
the maximum of the left, right, and intermediate HLL states.
Because the HLL state depends on the wave speeds themselves, we perform an
iteration procedure until the external waves could be moved out no further.
In the event that either HLLC or HLLD solver fails to produce a positive
pressure in any of the intermediate states, we fall back to the very robust
but dissipative HLLE Riemann solver \citep{einfeldt91} with a single
intermediate state $\mathbf{U}^*$.

\begin{figure}
\epsscale{0.8}
\plotone{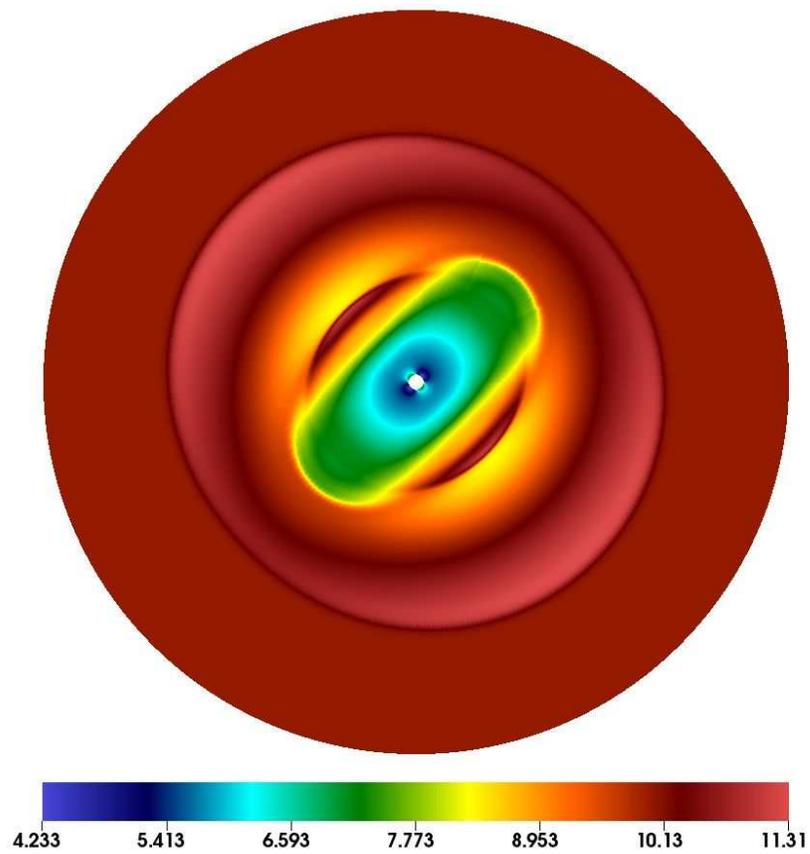}
\caption{Magnetic field magnitude at time $t=0.07$ from the blast wave problem.}
\label{fig_blast}
\end{figure}

The methods described above are used without modification with the source term
divergence cleaning algorithm (Eq. 6).
However, the GLM method introduces two additional waves moving with the speeds
$\pm c_h$ that carry changes in $B_n$ and $\psi$ only.
Because $c_h$ is the fastest wave speed, these waves bound the ``base'' Riemann
fan, comprised of 2, 3, or 5 waves in the HLLE, HLLC, and HLLD solvers,
respectively.
The intermediate states are readily obtained from the Rankine-Hugoniot
conditions at the bounding waves as
\begin{eqnarray}
B_n^*=\frac{B_{n,l}+B_{n,r}}{2}-\frac{\psi_r-\psi_l}{2c_h}, \nonumber \\
\psi^*=\frac{\psi_l+\psi_r}{2}-\frac{c_h(B_{n,r}-B_{n,l})}{2}.
\end{eqnarray}
These intermediate states serve as both the right and the left states for the
actual Riemann solver.
The advantage of this approach is that any possible jump in the normal
component of $\mathbf{B}$ is taken up by these additional external waves.

Very few genuinely three-dimensional test problems are available for spherical
grids.
For code verification we used a 3D blast wave problem similar to those presented
by \citet{gardiner08} and \citet{balsara09}.
The simulation region is constrained between $r_\mathrm{min}=0.01$ and
$r_\mathrm{max}=0.5$.
The radial cell width $\Delta r$ increased outward monotonically from 0.00052 to
0.047.
The inner boundary was treated as a perfectly conducting sphere with reflecting
boundary conditions imposed.
The initial conditions are $\rho=1$, $\mathbf{u}=0$, and $p=10\;(r<0.1)$,
$p=0.1\;(r>0.1)$.
The initial magnetic field is given by the standard potential solution for a
perfectly conducting sphere in a uniform external field, namely
\begin{eqnarray}
B_r=B_0\left(1-\frac{r_\mathrm{min}^3}{r^3}\right)\cos\theta, \\
B_\theta=-B_0\left(1+\frac{r_\mathrm{min}^3}{2r^3}\right)\sin\theta.
\end{eqnarray}
This solution was rotated such that the external field pointed in the direction
$(1/\sqrt{3},1/\sqrt{3},1/\sqrt{3})$.
We used $B_0=10$ and $\gamma=1.4$.
The system was evolved until $t=0.07$.

For this problem we chose a level 6 grid with 256 cells in the radial direction.
The GLM version of the numerical scheme was used with the HLLC Riemann solver
and WENO reconstruction.
Figure \ref{fig_blast} shows the magnitude of magnetic field at the end of the
simulation on a linear scale.
The flow structure of the solution is qualitatively similar to
\citet{gardiner08} and \citet{balsara09}, consisting of an outermost fast shock
wave and two dense shells of material elongated along the magnetic field.
This problem did not trigger the slope flattening or positivity correction
routines meaning it is not a very good stress test of the code.
Several more difficult problems simulating actual astrophysical plasma flows are
discussed next.

\section{Numerical solutions of test problems}
To illustrate the capabilities of the new model we present results from three
different simulations of solar system plasma environments.
The first is a dynamic MHD simulation of compressive structures in the solar
wind known as corotating interaction regions (CIRs).
This is a simple test problem with a strong degree of spherical symmetry.
The second is a simulation of the structure of the global heliosphere, including
regions on each side of the interface between the solar wind and LISM known as
the heliopause.
This problem involves mode complex transonic flows and a population of neutral
atoms in addition to the plasma.
Finally, our third test problem is a stationary structure of the Earth's
magnetosphere.
Unlike the two previous cases, this one is an example of a highly magnetized
plasma environment.

\subsection{Test problem 1: Corotating interaction regions in the solar wind}
Corotating interaction regions (CIRs) are compressive structures produced
through an interaction between high and low speed streams in the solar wind.
CIRs are fully formed by the time they reach Earth's orbit \citep{siscoe72,
gosling72}.
When the streams emanating from the Sun are approximately steady in the
co-rotating frame, these compression regions form spirals in the solar
equatorial plane that co-rotate with the Sun.
The leading edge of a CIR is a forward compressional wave propagating into the
slower solar wind ahead, whereas the tailing edge is a reverse wave propagating
back into the trailing high speed stream.
At large heliospheric distances the waves steepen into forward and reverse
shocks.
The entire plasma structure is convected with the solar wind and plays an
important role in the dynamics of the heliosphere.

\begin{figure}
\epsscale{0.7}
\plotone{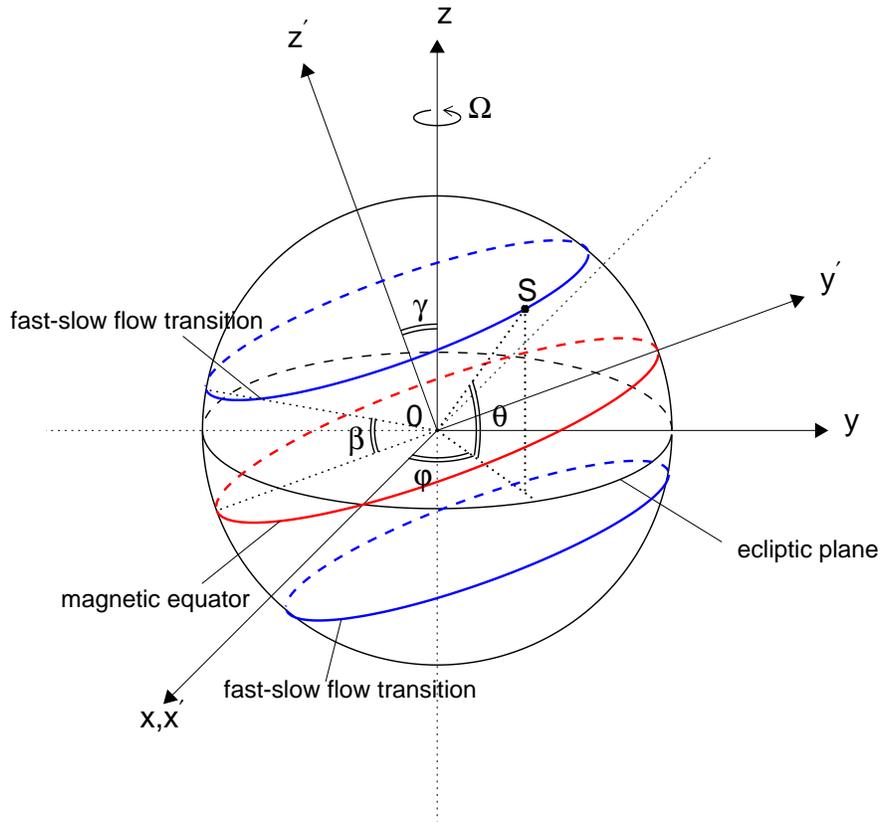}
\caption{A diagram of the assumed titled-dipole plasma flow geometry for the CIR
simulation.}
\label{fig_cirgeo}
\end{figure}

CIRs have been extensively studied using global MHD simulation \citep{pizzo94,
riley01, usmanov06}.
To generate CIRs in a global MHD simulation we adopt the tilted-dipole flow
geometry of \citet{pizzo82} at the inner boundary, which is illustrated in
Figure \ref{fig_cirgeo}.
In this figure $0xyz$ is the fixed (heliographic) frame, where $z$ is the solar
rotation axis, and $0x'y'z'$ is a frame aligned with the Sun's magnetic axis
$z'$.
The parameter $\gamma$ is the dipole tilt angle, and $\beta$ is the latitude
of the fast-slow transition boundaries (blue circles) in the coordinate system
$0x'y'z'$.
In the simulation discussed below we used $\beta=\pm30^\circ$.

Following \citet{pogorelov07}, one readily derives a quadratic equation for the
latitude of the transition line $\theta$ as a function of the azimuthal angle
$\varphi$,
\begin{equation}
a\sin^2\theta+b\sin\theta+c=0,
\end{equation}
where
\begin{eqnarray}
a&=&\cos^2\gamma+\sin^2\gamma\tan^2\varphi(\cot\gamma\cos\gamma+\sin\gamma)^2
\nonumber \\
b&=&2\sin\beta\cos\gamma(1+\tan^2\varphi) \\
c&=&\sin^2\beta(1 + \tan^2\varphi\cos^2\gamma)-\cos^2\beta\sin^2\gamma\tan^2\varphi.
\nonumber
\end{eqnarray}
Note that when $\beta=0$, $\theta(\varphi)$ reduces to the expression for the
latitude of the magnetic equator given by Eq. (A6) of \citet{pogorelov07}.

We simulated a region of the solar wind between $r_\mathrm{min}=0.5$ AU and
$r_\mathrm{max}=30$ AU using 512 concentric grid layers of variable thickness
(increasing outward).
At 1 AU we assume the following conditions: density $n=3.5$ cm$^{-3}$ and radial
velocity $u=800$ km/s in the fast solar wind and $n=7$ cm$^{-3}$, $u=400$ km/s
in the slow solar wind.
The radial component of the magnetic field at 1 AU was $B_r=28$ $\mu$G.
These conditions were extended to the inner boundary using the conventional
Parker solution for the solar wind and its magnetic field \citep{parker58}.
The dipole tilt angle was taken to be $\gamma=20^\circ$.
The boundary (shear layer) between the fast and the slow solar wind flows was
located at a latitude of $30^\circ$ in the coordinate system aligned with the
dipole axis.
This simulation was performed on a level 6 geodesic grid.
We chose the HLLC solver to evolve the time-dependent MHD equations, combined
with the GLM divergence cleaning method; WENO reconstructions was used in all
directions.

\begin{figure}
\epsscale{1.11}
\plottwo{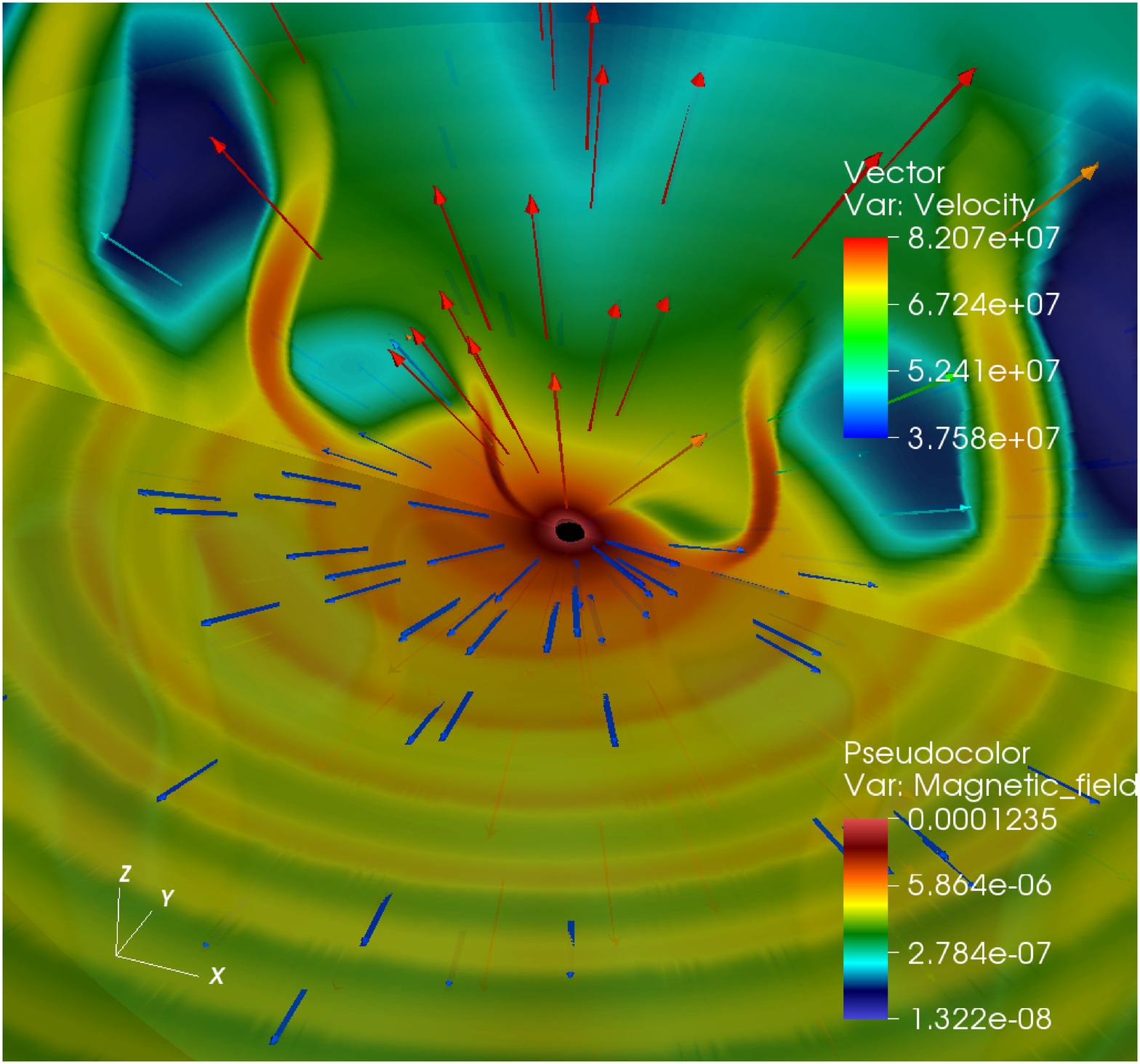}{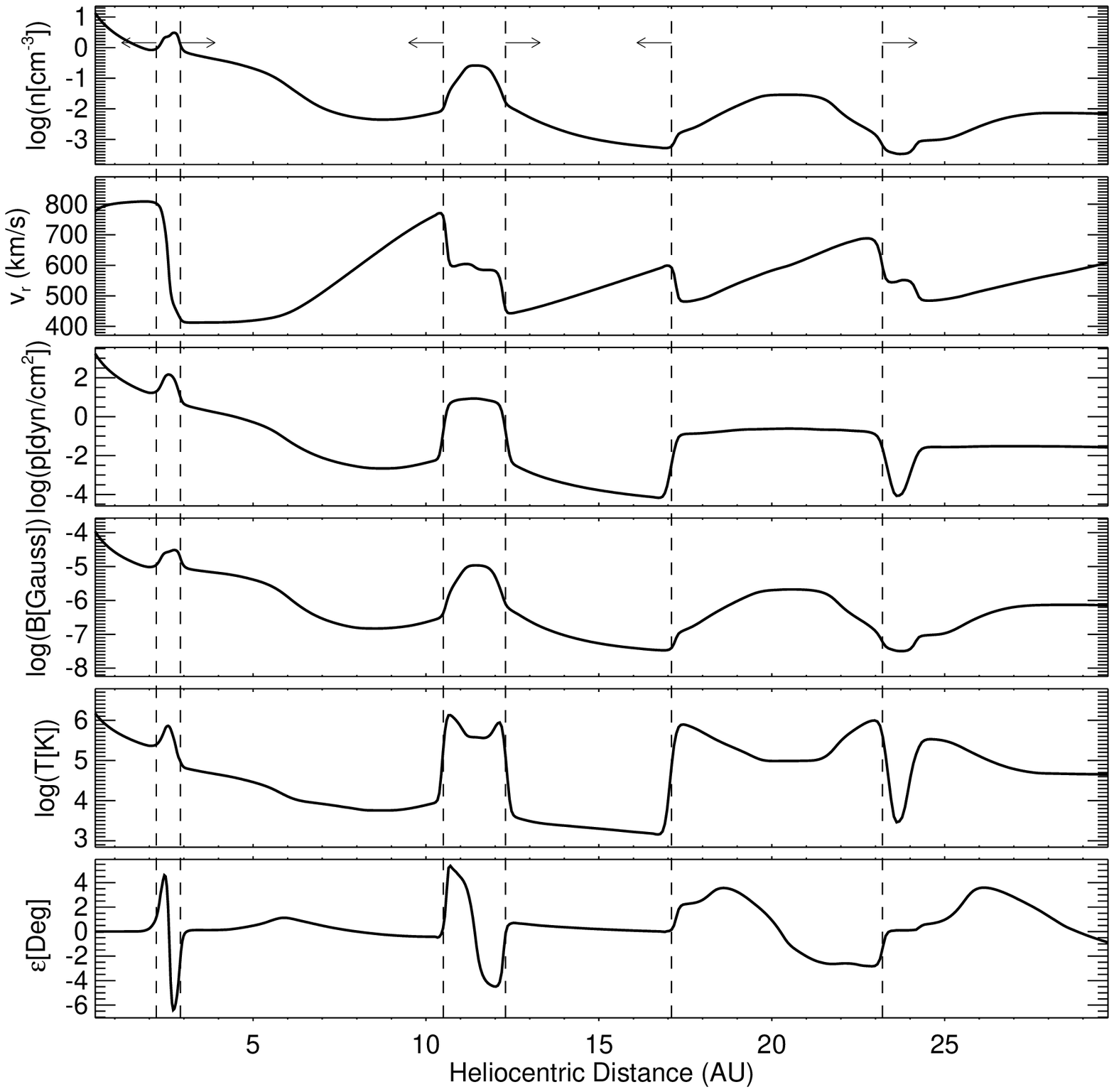}
\caption{Left: Magnetic field magnitude (log scale) in the meridional plane
($xz$) and the solar equatorial plane ($xy$) for the CIR simulation.
Arrows are the plasma velocity vectors.
Right: radial profiles along the direction ($\theta$, $\varphi$) = ($25^\circ$,
$135^\circ$) of (from top to bottom): plasma density (log scale), radial
velocity, log thermal pressure, log magnetic field intensity, log temperature,
and the north-south flow deflection angle $\epsilon$.
Arrows mark the forward (pointing right) and reverse (pointing left) propagation
of wave fronts.}
\label{fig_cir}
\end{figure}

Figure \ref{fig_cir} (left) shows the logarithm of the magnetic field magnitude
in the $xz$ and $xy$ planes using a cutout plot.
Plasma velocity vectors are shown as arrows of variable length.
The CIRs can be visually identified as higher density and magnetic field
intensity regions (red).
The maximum latitudinal extent of CIRs is given by the sum of the angle between
the rotation and the dipole axes and the extent of the slow solar wind in
the frame aligned with the dipole axis, i.e., $\gamma+\beta=50^\circ$.
In the equatorial plane, the spiral CIR structure is seen to be bounded by
shock-like discontinuities.

Several characteristic CIR features can be recognized in the plasma radial
profiles shown in Figure \ref{fig_cir} (right).
We chose the profile along the direction $25^\circ$ northern latitude relative
to the solar equatorial ($xy$) plane.
The forward-reverse shock pairs are commonly observed at mid-latitudes, below
the heliographic latitude of $26^\circ$ \citep{gosling99}.
They are shown with vertical dashed lines in the Figure.
Shock pairs associated with CIRs are believed to be responsible for the observed
26-day recurrent decreases in galactic cosmic-ray intensity \citep{kota91,
mckibben99}.
Other features, such as the south-north flows are also identified through the
north-south flow deflection angle $\epsilon=\sin^{-1}(-u_\theta/|\mathbf{u}|)$
shown in the bottom panel.
The transitions from northward (positive) to southward (negative) velocity are
separated by roughly one Carrington rotation period (26 days) in our simulation.
We conclude that the model is capable of reproducing the essential CIR features
and is consistent with the earlier simulations of this phenomenon.

\subsection{Test problem 2: The global heliosphere}
The energy density in a supersonic stellar wind, such as the solar wind,
decreases in inverse proportion to the square of the distance from the star.
Eventually the outflow is unable to maintain pressure balance with the
galactic environment near the star, comprised mostly of partially ionized
hydrogen gas.
The stellar wind undergoes a transition to a subsonic flow at a structure called
a termination shock.
A tangential discontinuity called an astropause (heliopause for the solar wind)
separates the shocked stellar flow from the interstellar gas.
A bow shock may develop in front of the astropause if the relative motion
between the star and LISM is supersonic.
In the case of heliosphere, the region between the termination shock and the
boundary is called the heliosheath.
The theory of stellar wind interfaces (as applied primarily to the heliosphere)
has been developed in \citet{parker61}, \citet{axford72}, and \citet{baranov76}.
Recent three-dimensional MHD simulations of the interface could be found in
\citet{pogorelov07} and \citet{opher07}.

\begin{figure}
\epsscale{0.7}
\plotone{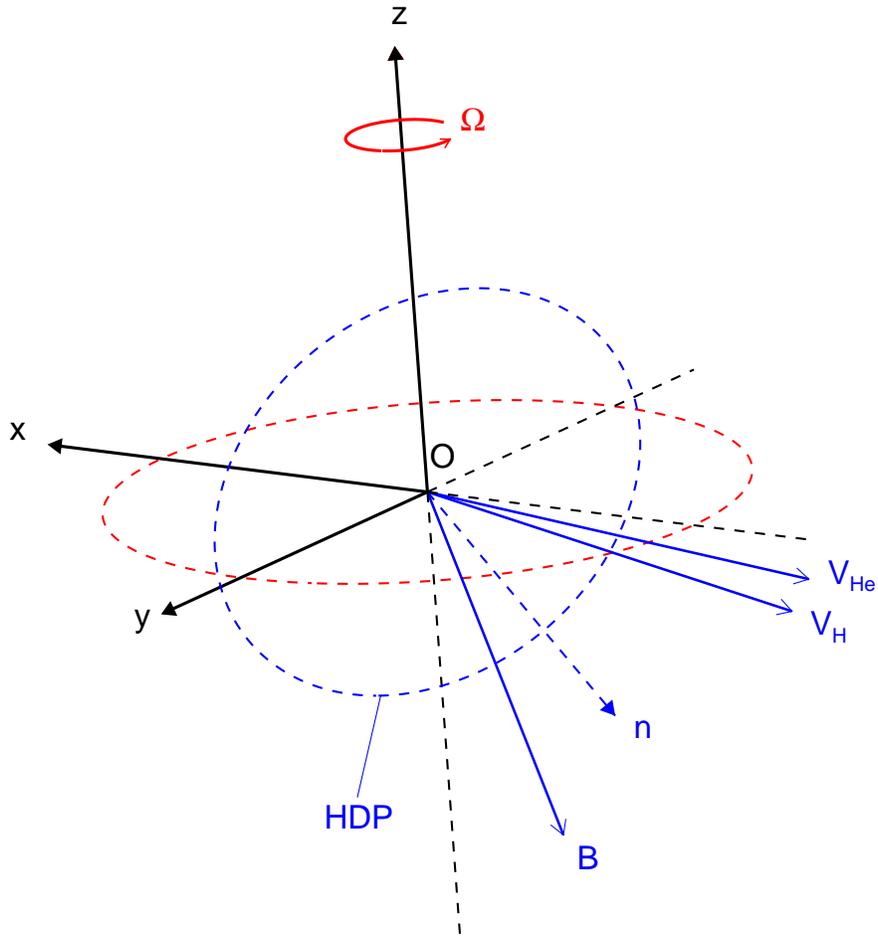}
\caption{The heliographic coordinate system used in the simulation of the global
heliosphere.
The directions of the flow of interstellar hydrogen ($\mathbf{V}_\mathrm{H}$)
and helium ($\mathbf{V}_\mathrm{He}$) span the so-called hydrogen deflection
plane (HDP) with a normal $\mathbf{n}$.
Here the interstellar magnetic field $\mathbf{B}$ lies in the HDP, with an angle
of $45^\circ$ relative to $\mathbf V_{He}$.}
\label{fig_hdp}
\end{figure}

To simulate the structure of the heliospheric interface we used a relatively
coarse level 5 geodesic grid with 240 radial points.
As in the CIR problem, the concentric layer spacing was nonuniform with the
smallest cells at the inner radial boundary located at 10 AU; the outer boundary
was placed at 900 AU.
A heliographic coordinate system is used here, where the $z$ axis is aligned
with the solar rotation axis \citep{beck05}, and the $x$ axis is in the plane
formed by the $z$ axis and the interstellar helium flow direction
\citep{lallement05}.
The $y$ axis completes the right-handed orthogonal system.
The geometry of the problem is illustrated in Figure \ref{fig_hdp}.

The heliospheric configuration computed here is representative of a solar
minimum \citep{florinski11}.
At 1 AU we assume the following conditions: density $n=3.5$ cm$^{-3}$ and radial
velocity $u=800$ km/s at heliographic latitudes above $30^\circ$ (fast solar
wind) and $n=7$ cm$^{-3}$, $u=400$ km/s at low latitudes (slow solar wind).
The magnetic field is a Parker spiral with a radial component $B_r=28$ $\mu$G at
1 AU.
The azimuthal magnetic field component is a function of the solar wind speed.
The heliospheric current sheet is not included in this simulation, so that the
solar magnetic field is always directed outward from the Sun.
The observed current sheet is between $10^4$ km \citep[1 AU,][]{winterhalter94}
and a few times $10^5$ km \citep[heliosheath,][]{burlaga11} in width, which is
much too narrow to be resolved with a global model.

The interstellar flow has a total density of 0.2 cm$^{-3}$, and is ionization
rate of 0.25.
Its velocity vector is $\mathbf{V}_\mathrm{He}=(-26.3,\,0,\,-0.23)$ km/s in the
chosen heliographic coordinate system.
The interstellar magnetic field lies in the so-called hydrogen deflection plane
(the plane spanned by the velocity vectors of neutral interstellar hydrogen and
helium) and is inclined by $45^\circ$ with respect to the LISM flow vector.
Its components are $(-1.3,\,1.38,\,-2.32)$ $\mu$G in our coordinate system.
The temperature of both ionized and neutral components in the LISM is taken to
be 6530 K.
The neutral and the plasma fluids are coupled via the charge exchange process
\citep{axford72}.
We simulate both fluids using the same code by explicitly fixing $\mathbf{B}=0$
for the neutral hydrogen.
The charge exchange terms used are those of \citet{pauls95}.
For simplicity we only include interstellar hydrogen in this simulation and
ignore atoms produced by charge exchange in the heliosheath or the solar wind.
To separate the interstellar region from the heliosphere we use a passively
advected indicator variable $q$ which satisfies the equation
\begin{equation}
\frac{\partial(\rho q)}{\partial t}+\nabla\cdot(\rho q\mathbf{u})=0.
\end{equation}
The indicator variable is set to 1 in the solar wind and $-1$ in the
interstellar flow.
The condition $q=0$ then gives the location of the heliopause.

\begin{figure}
\epsscale{1.11}
\plottwo{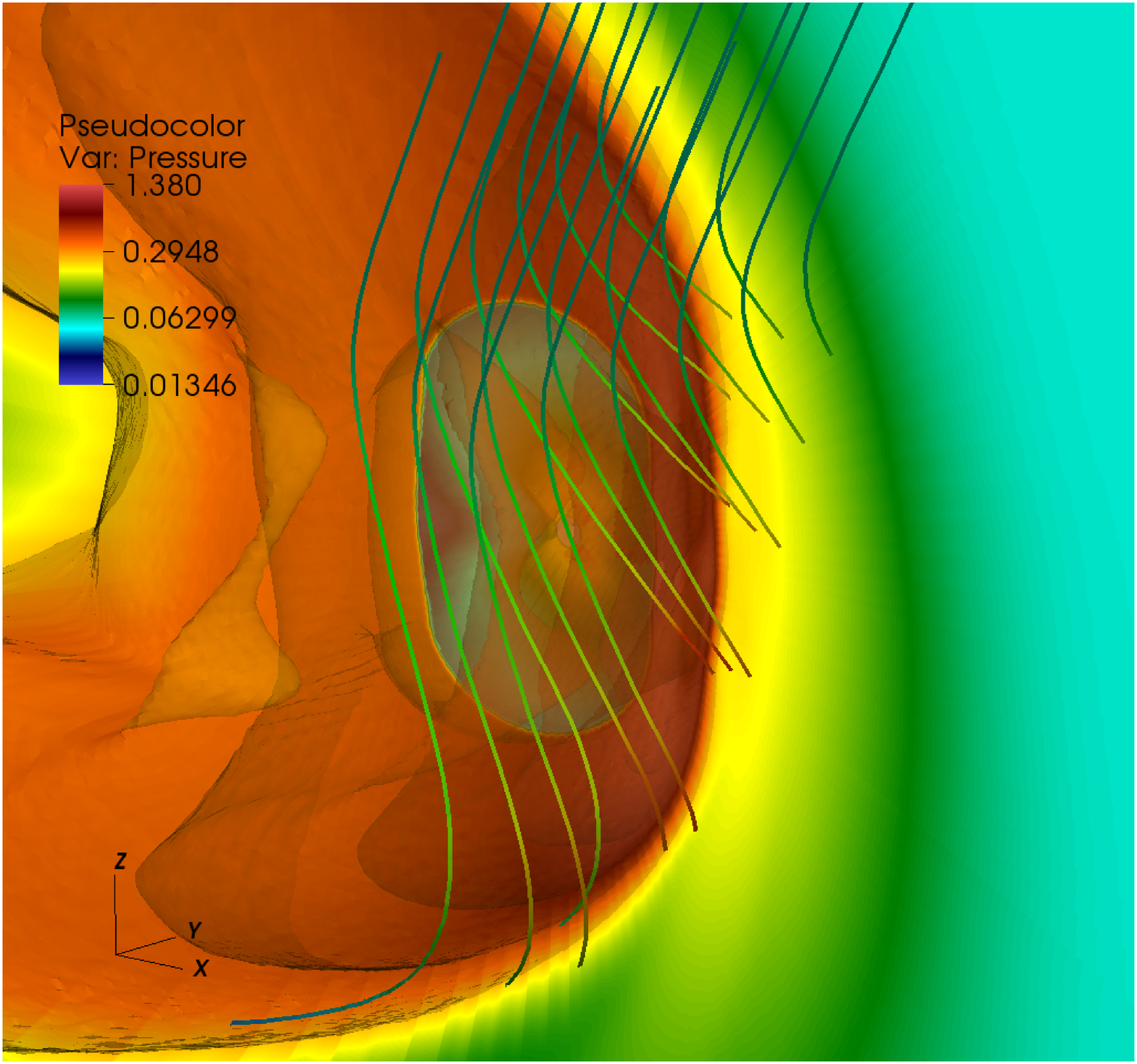}{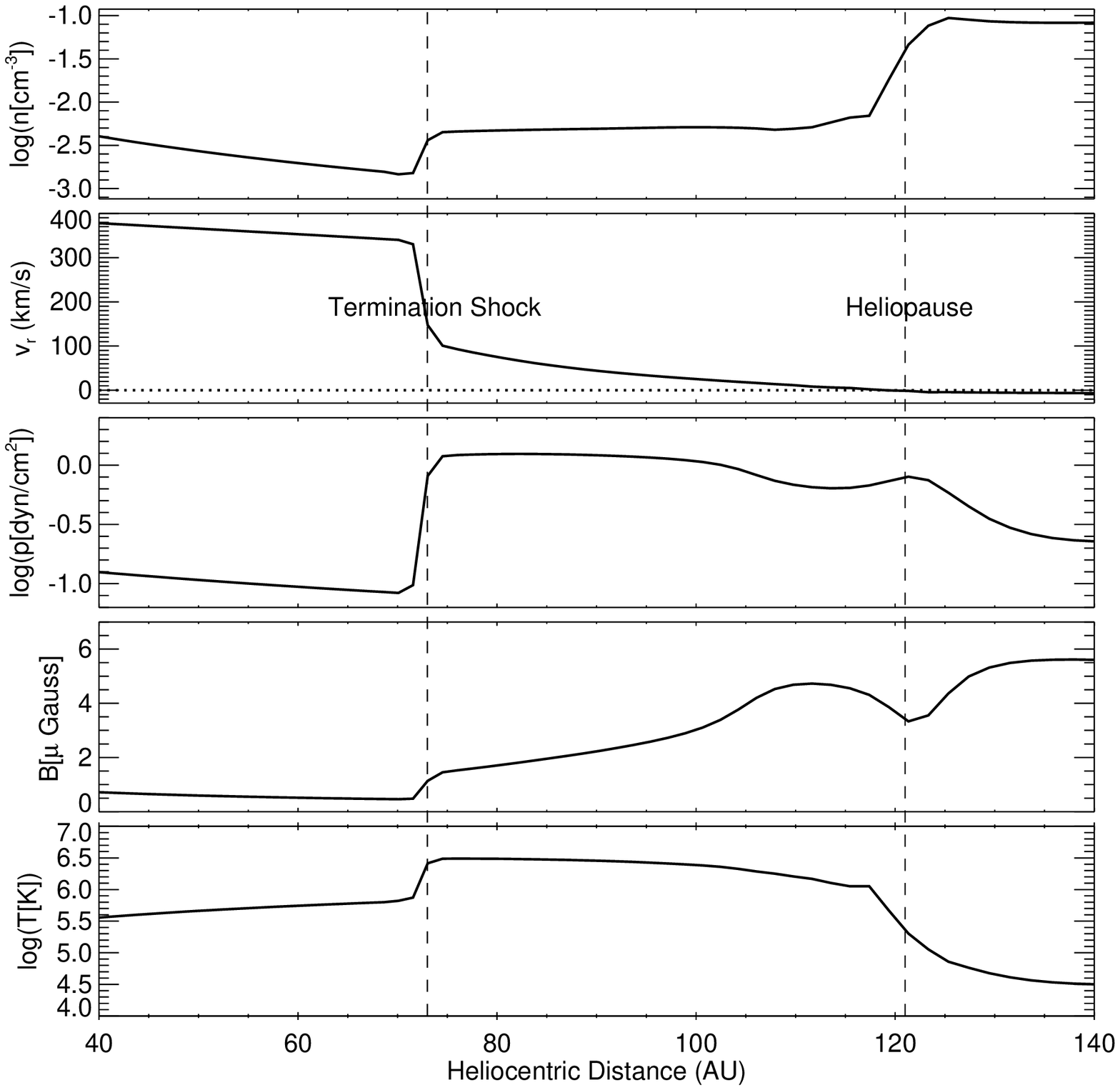}
\caption{Left: Constant plasma pressure surfaces (log scale) cut by the
meridional ($xz$) plane for the heliosphere simulation.
Selected magnetic field lines in the LISM are shown.
Right: Radial profiles in the upwind (nose) direction of (from top to bottom):
log plasma number density, radial speed, log thermal pressure, magnetic field
magnitude, and log temperature.
The positions of the termination shock and the heliopause are marked with
vertical dashed lines.}
\label{fig_helio}
\end{figure}

We chose the HLLC solver for this work because of its more robust handling of a
strong flow shear between the fast and the slow solar wind.
We used the GLM $\nabla\cdot\mathbf{B}$ control method and WENO reconstruction
in all directions.
Simulations were run until a steady state was achieved which took about 300
years of simulated time.
Figure \ref{fig_helio}, left, shows a cutout view of the heliospheric interface.
Surfaces of constant plasma pressure are plotted together with magnetic field
lines in the LISM, illustrating their draping around the heliopause (the
transition between the red and the green colors).
The innermost pressure surface approximately traces the outline of the
termination shock.

We show radial profiles of several physical quantities in the upwind, or
``nose'' direction in the right panel of Figure \ref{fig_helio}.
Before the termination shock, located at 67 AU in this simulation, the solar
wind velocity is gradually decreasing because of a loss of momentum to charge
exchange with interstellar hydrogen.
In the heliosheath, the plasma density is nearly a constant while the magnetic
pressure increases toward the heliopause where the flow becomes essentially
stagnant.
The effective heliosheath temperature ($\sim 3\times 10^6$ K) is that of the
solar-wind and pickup-ion mixture, which is significantly higher than that of
the core solar wind \citep[$\sim 2\times 10^5$ K,][]{richardson08}.
From the top panel one can see that the density on the interstellar side of the
heliopause is some 25 times higher than in the heliosheath.
There is a very weak bow shock in this model barely visible in the pressure and
temperature profiles.

The results presented here were obtained using a single population of neutral
hydrogen (the interstellar atoms).
The computer code is actually capable of integrating conservation laws for
multiple neutral hydrogen populations.
It would be straightforward to include the heliosheath energetic neutral atoms
and the neutral solar wind atoms in a simulation, at an added computational
time expense \citep[e.g.,][]{williams97}.

\subsection{Test problem 3: Magnetosphere of Earth}
The Earth's magnetosphere is a product of an interaction between the supersonic
solar wind and the geomagnetic field.
Two major discontinuities, the bow shock and the magnetopause, are located
between the undisturbed solar wind region and the geomagnetic field.
The magnetosheath, filled with shocked solar wind plasma, lies between the bow
shock and the magnetopause, which is the external boundary of the magnetosphere.
The magnetopause thus separates the hot, tenuous magnetospheric plasma from the
cold and dense solar wind plasma in the magnetosheath.
Global MHD simulations, coupled with ionospheric models, have been widely used
to study large-scale processes in the magnetosphere \citep[e.g.,][]{fedder95,
tanaka95, raeder99, hu07}.

The geomagnetic field can be treated as a dipole field in the inner
magnetosphere, its strength varying as $r^{-3}$, where $r$ is the distance from
the center of the Earth.
The thermal pressure varies more modestly leading to a very low plasma $\beta$
(the ratio of the plasma thermal pressure to the magnetic field pressure) in the
inner magnetosphere.
Such low values of $\beta$ ($\sim 10^{-5}-10^{-4}$) tend to produce numerical
errors with conservative numerical schemes \citep{raeder99}.
To overcome this difficulty, the dipole field is treated apart from the total
magnetic field  according to the decomposition method introduced by
\citet{tanaka95}.
The momentum and energy fluxes in the Riemann solvers are revised accordingly.
The WENO reconstruction method is used in all directions and the GLM algorithm
is used to control $\nabla\cdot{\mathbf B}$.
An interested reader will find more details on the GLM-MHD equations with a
dipole field decomposition in the Appendix.

The Geocentric Solar Magnetospheric (GSM) coordinate system is used in this
simulation.
It is centered at Earth, and the $x$, $y$, and $z$ axes point to the Sun, the
dawn-dusk direction, and along the north dipole axis, respectively.
We choose the inner boundary to be a sphere with a radius $r=3 R_\mathrm{E}$
(Earth radii), and apply the Dirichlet boundary conditions.
In particular, the number density is 370 cm$^{-3}$, which is 1/27 of a typical
value in the ionosphere.
The thermal pressure is $4.65\times10^{-10}$ dyn/cm$^2$, which is 9 times
smaller than its ionospheric value.
The magnetic field is taken to be a dipole field at the inner boundary.
For the sake of simplicity, the magnetosphere-ionosphere electrostatic coupling
\citep[e.g.,][]{janhunen98} is not included, therefore the feedback of the
ionosphere on the magnetosphere is ignored.
We simply set the velocity to zero, which means there is no convection at
the inner boundary.
The free outer boundary is located at $r=100 R_E$.

We simulate a common configuration with a southward interplanetary magnetic
field (IMF) of 50 $\mu$G.
The solar wind velocity is 600 km/s along the Sun-Earth line (the negative $x$
direction), its number density is 5 cm$^{-3}$ and temperature
$9.1\times 10^4$ K.
The magnetic field is initially calculated as a superposition of a dipole field,
centered at the origin, and a mirror dipole, located at
$(30 R_\mathrm{E},\,0,\,0)$,
The field on the sunward side is subsequently replaced with the solar wind
field with $B_z=-50$ $\mu$G to make the initial configuration divergence free.
In the simulation we used a level 6 geodesic grid and 256 grid points along the
radial direction.

\begin{figure}
\epsscale{1.11}
\hspace{-5mm}
\plottwo{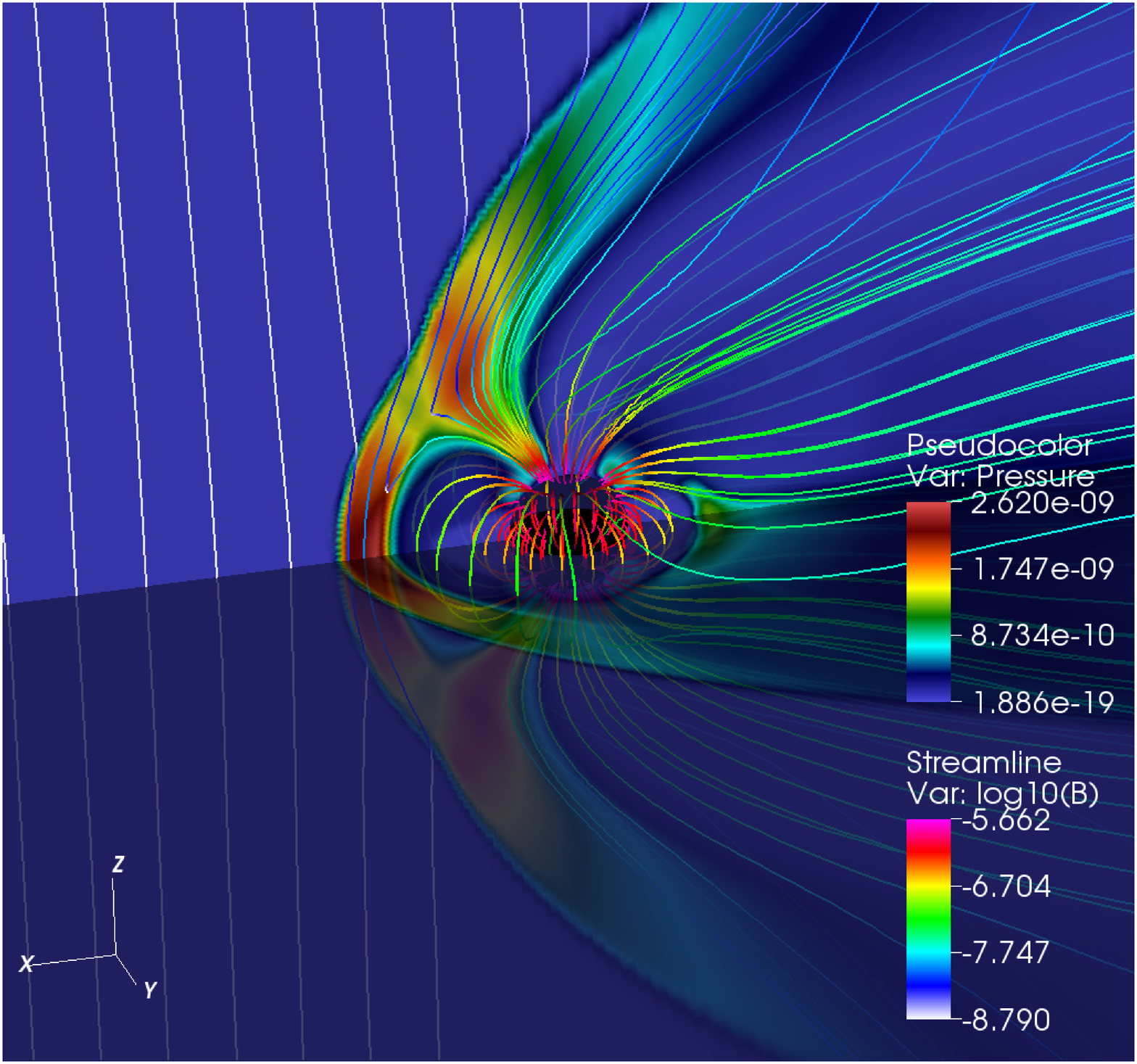}{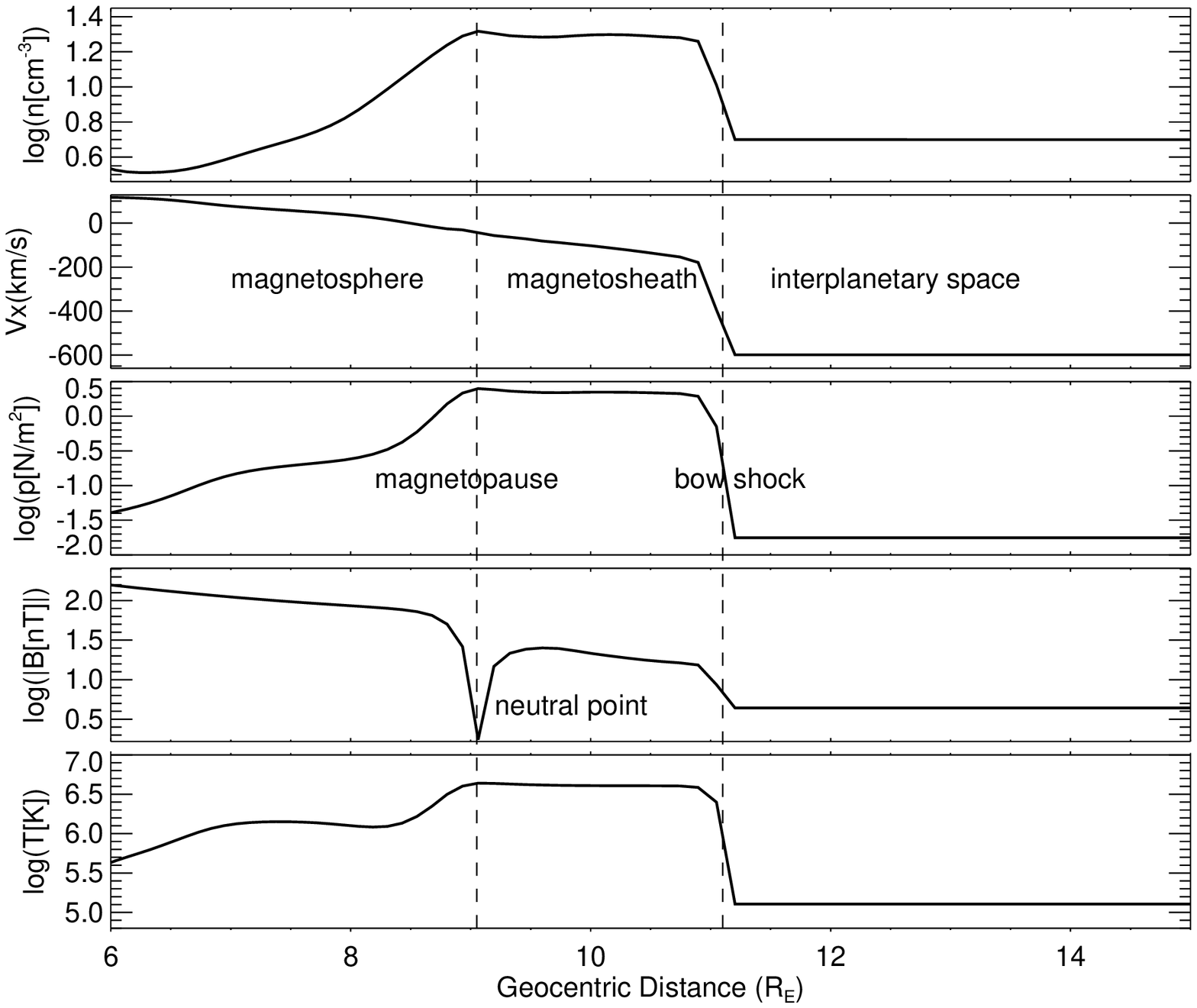}
\caption{Left: plasma pressure (color) and magnetic field lines for the
magnetosphere simulation.
Plane cuts for $z=0$ and $y=0$ are shown.
The magnitude of the magnetic field is shown by the color of the field lines in
the figure.
Right: radial profiles along the Sun-Earth line of (from top to bottom):
log plasma number density, velocity, log thermal pressure, log magnetic field
intensity, and log temperature.
The bow shock and the magnetopause are marked by vertical dashed lines.}
\label{fig_mag}
\end{figure}

A steady state configuration is obtained some 30 minutes (simulated time) into
the simulation.
The left panel of Figure \ref{fig_mag} shows the color contours of the thermal
pressure in the meridional ($xz$) plane and in the equatorial ($xy$) plane.
The geomagnetic field and the IMF lines of force are also plotted.
In the equatorial plane the geomagnetic field points northward, being opposite
to the polarity of the southward IMF.
We can see that the magnetosphere is open to the interplanetary medium and the
geomagnetic field lines connect with the IMF \citep{dungey61}.
In that case the solar wind plasma momentum and energy can be transported into
the magnetosphere through the site of magnetic reconnection.
We did not observe surface waves or vortices induced by the Kelvin-Helmholtz
instability \citep[e.g.,][]{guo10} along the low-latitude magnetopause (the
surface of the magnetopause is smooth in the equatorial plane).

The profiles of the physical quantities along the Sun-Earth line are shown in
the right panel of Figure \ref{fig_mag}.
The magnetopause is located at the neutral point for southward IMF case.
The $x$ velocity component approaches zero at the subsolar point, where the
Sun-Earth line intersects the magnetopause.
The shocked plasma becomes dense and hot in the magnetosheath, compared with the
undisturbed solar wind.
For southward IMF, the neutral point is found from the magnetic field strength
profile (fourth panel from the top), where magnetic reconnection could occur in
the presence of dissipation.

Our result has all the relevant features of a typical MHD magnetospheric
simulation.
In this illustrative solution, we only calculate a steady state representative
magnetosphere.
Of course, the model can be also used with more realistic time-dependent IMF
conditions derived from observations.

\section{Summary}
In this report we have presented a novel approach to numerical modeling of
space plasma flows using geodesic spherical meshes with a nearly uniform solid
angle coverage.
This approach avoids the singularity on the symmetry axis inherent in polar
spherical grids, leading to improved efficiency by allowing larger time steps.
Our integration technique for gas-dynamic or MHD conservation laws is based on
dimensionally unsplit time advance and uses two-dimensional reconstruction on
the surface of a sphere.
The new code has a number of useful features, such as a choice of multiple
nonlinear Riemann solvers, weighted reconstruction limiters, and slope
flattening to reduce possible oscillations near strong shocks.

We have tested the new model on several common problems in space physics: a
formation of corotating interaction regions in the solar wind, global modeling
of the heliospheric interface, and finally, the magnetosphere of a planet.
Our results are consistent with those found in the literature and every feature
of the resulting structures is well reproduced.
At this time the model lacks an adaptive mesh refinement feature, which would
permit a superior numerical resolution of shocks and discontinuities.
Whereas a hexagonal (Voronoi) grid cannot be easily refined, its dual Delaunay
grid can.
The process starts with the original icosahedron that divides a sphere into 20
identical spherical triangles.
Each triangle then may be recursively subdivided into four smaller triangles by
connecting the midpoints of the original cell edges with great circle arcs.
The Delaunay mesh is therefore naturally amenable to refinement based on an
oct-tree formulation.
Because each locally refined zone is further split in the radial direction,
this is tantamount to each 3D patch giving rise to 8 identically-sized refined
patches if it is to undergo one more level of refinement. 

The model could be potentially adapted to solve problems where the compact
object is not at the center of the region of interest.
For example, following \citet{tanaka00}, one could introduce a non-concentric
grid, where different spherical layer boundaries are offset from the origin.
The offset distance increases for each subsequent layer, so that the mesh
becomes denser in one direction and more rarefied in the opposite direction.
Such an arrangement could be more efficient for modeling, e.g., a
magnetosphere with a long tail.

The new code by itself could be a valuable tool to investigate plasma flows
around a source whose dimensions are small compared with the scale of the flow.
Nevertheless, its chief intended purpose is to provide plasma background for
subsequent simulations of the transport of energetic charged particles
in the solar system and other astrophysical environments.
Additional modules, recently added to the code, calculate the diffusion
coefficients and drift velocity vectors based on magnetic field and other plasma
properties.
The use of geodesic grids will permit a more efficient calculation of phase
space trajectories in the stochastic integration method popular in cosmic-ray
transport work \citep{ball05, florinski09}.
The difference with polar grid-based models is expected to be quite pronounced
in the polar regions of the heliosphere, where the diffusion and drift
coefficients are typically very large.

\acknowledgments
V.F. and X.G. were supported, in part, by NASA grants NNX10AE46G and NNX12AH44G,
NSF grant AGS-0955700, and by a cooperative agreement with NASA Marshall Space
Flight Center NNM11AA01A.

\appendix

\section{Dipole field decomposition}

In the magnetosphere, the external field $\mathbf B_1=\mathbf B-\mathbf B_d$, 
where $\mathbf B$ is the total magnetic field, and $\mathbf B_d$ is the
internal dipole field.
Since $\mathbf B_d$ is both curl-free (no current) and divergence free, we can
write
\begin{equation} 
\nabla\cdot\left(\mathbf B_d\mathbf B_d-\frac{1}{2}B_d^2{\mathbf I}\right)=0,
\end{equation}
\begin{equation} 
(\nabla\times{\mathbf B_d})\cdot(\mathbf u\times\mathbf B)=0.
\end{equation}
Using (A1) the momemtum flux from Eq. (4) can be expressed as
\begin{equation}
\rho{\mathbf{uu}}+p{\mathbf I}
-\frac{1}{4\pi}\left(\mathbf{BB}+\frac{1}{2}B_d^2{\mathbf I}
-\mathbf B_d\mathbf B_d\right).
\end{equation}
Next, from (A2) we obtain
\begin{equation} 
{\mathbf B_d}\cdot\nabla\times(\mathbf u\times\mathbf B)
-\nabla\cdot(\mathbf u\times\mathbf B)\times{\mathbf B_d}=0,
\end{equation}
which, upon substitution into the magnetic induction equation
\begin{equation} 
\frac{\partial\mathbf B}{\partial t}
-\nabla\times(\mathbf u\times\mathbf B)=0
\end{equation}
yields
\begin{equation} 
\mathbf B_d\cdot\frac{\partial\mathbf B_1}{\partial t}
-\nabla\cdot[\mathbf B(\mathbf u\cdot\mathbf B_d)
-\mathbf u(\mathbf B\cdot\mathbf B_d)]=0.
\end{equation}
We now define
\begin{equation}
p_1^*=p_g+\frac{B_1^2}{8\pi},\quad
e_1=\frac{\rho u^2}{2}+\frac{p_g}{\gamma-1}+\frac{B_1^2}{8\pi}.
\end{equation}
Using these definitions the energy equation may be written as
\begin{eqnarray}
\frac{\partial e_1}{\partial t}
+\frac{\mathbf B_d}{4\pi}\cdot\frac{\partial\mathbf B_1}{\partial t}
+\nabla\cdot\left\lbrace(e_1 + p_1^*){\mathbf u}\right. \nonumber \\
\left.+\frac{1}{4\pi}
[\mathbf u(\mathbf B_1\cdot\mathbf B_d)+\mathbf u(\mathbf B\cdot\mathbf B_d)
-\mathbf B(\mathbf u\cdot\mathbf B_1)-\mathbf B(\mathbf u\cdot\mathbf B_d)]
\right\rbrace=0.
\end{eqnarray}
Combining equations (A6) and (A8) yields
\begin{equation}
\frac{\partial e_1}{\partial t}
+\nabla\cdot\left\lbrace(e_1 + p_1^*){\mathbf u}
-\frac{1}{4\pi}[\mathbf B_1(\mathbf u\cdot\mathbf B_1)
-\mathbf u(\mathbf B_d\cdot\mathbf B_1)+ 
\mathbf B_d(\mathbf u\cdot\mathbf B_1)]\right\rbrace=0.
\end{equation}

Using (A3) and (A9) the system of GLM-MHD equations with dipole field
decomposition may be written as
\begin{equation}
\frac{\partial\rho}{\partial t}+\nabla\cdot(\rho{\mathbf u})=0,
\end{equation}
\begin{equation}
\frac{\partial(\rho{\mathbf u})}{\partial t}+\nabla\cdot
\left[\rho{\mathbf{uu}}+p^{*}{\mathbf I}-\frac{1}{4\pi}
\left({\mathbf{BB}}+\frac{1}{2}B_d^2{\mathbf I}-\mathbf{B}_d\mathbf{B}_d\right)
\right]=0,
\end{equation}
\begin{equation}
\frac{\partial{\mathbf B_1}}{\partial t}+\nabla\cdot({\mathbf{uB}-\mathbf{Bu} 
+\psi{\mathbf I}})=0,
\end{equation}
\begin{equation}
\frac{\partial e_1}{\partial t}+\nabla\cdot\left\lbrace(e_1+p_1^{*}){\mathbf u}
-\frac{1}{4\pi}\left[(\mathbf{u}\cdot\mathbf B_1)\mathbf B_1
-(\mathbf B_d\times\mathbf u)\times\mathbf B_1\right]\right\rbrace=0,
\end{equation}
\begin{equation}
\frac{\partial\psi}{\partial t}+c_h^2\nabla\cdot\mathbf{B}_1
=-\frac{c_h^2}{c_p^2}\psi,
\end{equation}
Note that the system (A10)--A(14) uses $\mathbf B_1$ and $e_1$ instead of
$\mathbf B$ and $e$ as conserved quantities

Consider the simplest three-state HLL solver \citep{harten83}.
Its Riemann flux is given by
\begin{equation}
\mathbf{F}=\left\lbrace\begin{array}{ll}
\mathbf{F}_l, & S_l>0, \\
\mathbf{F}_{lr}, & S_l\leq 0\leq S_r, \\
\mathbf{F}_r, & S_r<0,
\end{array}\right.
\end{equation}
where $\mathbf{F}_l=\mathbf{F}(\mathbf{U}_l)$ and
$\mathbf{F}_r=\mathbf{F}(\mathbf{U}_r)$ are the left and right unperturbed
fluxes, respectively.
The intermediate flux $\mathbf{F}_{lr}$ is given by
\begin{equation}
\mathbf{F}_{lr}=\frac{S_r\mathbf{F}_l-S_l\mathbf{F}_r
+S_l S_r(\mathbf{U}_r-\mathbf{U}_l)}{S_r - S_l}.
\end{equation}
Since only the definition of a conserved flux is required to solve (A15),
the system (A10)--(A14) can be readily used in place of (4).

The decomposition of magnetic field does not affect the GLM scheme.
For example, in the $x$ direction we have two GLM equations,
\begin{equation}
\frac{\partial B_{1x}}{\partial t}+\frac{\partial\psi}{\partial x}=0,
\end{equation}
\begin{equation}
\frac{\partial\psi}{\partial t}+\frac{\partial(c_h^2 B_{1x})}{\partial x}
=-\frac{c_h^2}{c_p^2}\psi.
\end{equation}
One can see that the external field component $B_{1x}$ can be integrated
directly because the internal field ($\mathbf B_d$ related terms) does not
appear in these equations. 


\end{document}